\documentclass[showpacs,preprintnumbers,amsmath,amssymb]{revtex4}
\usepackage{graphicx}
\usepackage{amstext}
\usepackage{amsmath}
\usepackage{latexsym}
\usepackage{amsfonts}
\usepackage{bm}
\usepackage{amssymb}
\usepackage{amscd}
\usepackage{slashbox}

\usepackage{array,multirow}

\begin{document}
\setcounter{page}{1}
\title{Multipartite Quantum Secret Sharing using a Single Not-So-Weak Qubit}

\author{Yun Jin Choi}

\affiliation{Department of Physics and Center for Quantum Spacetime,
Sogang University, Seoul 121-742, Korea}

\author{Young-Jai Park}
\affiliation{Department of Physics and Center for Quantum Spacetime,
Sogang University, Seoul 121-742, Korea}

\author{Chil-Min Kim}
\affiliation{National Creative Research Initiative Center for
  Controlling Optical Chaos, Pai-Chai University, Daejeon 302-735, Korea}

\author{Jaewan Kim}
\affiliation{School of Computational Sciences, Korea Institute for
Advanced Study, Seoul 130-012, Korea}

\date{\today}

\begin{abstract}
  We propose a new quantum secret sharing scheme using a single
 non-entangled qubit. In the scheme, by transmitting a qubit to the next
 party sequentially, a sender can securely transmit a secret message
 to $N$ receivers who could only decode the message cooperatively after
 randomly shuffling the polarization of the qubit. We explain this quantum
 secret sharing scheme into the one between a sender and two receivers,
 and generalize the scheme between a sender and $N$ receivers. Since
 our scheme is capable of using a faint coherent pulse as a qubit, it is
 experimentally feasible within current technology.
\end{abstract}



\maketitle
\section{Introduction}
  A cryptography based on quantum mechanics has received much attention
 since the seminal work on quantum key distribution (QKD) by Bennett and
 Brassard (BB84) \cite{k1} and Ekert (E91) \cite{k4}. In this scheme,
 information can be securely transmitted to a privileged person in novel
 ways \cite{k2,k3}. In distinction to this type of one-to-one communication,
 recently, another quantum cryptography scheme, named quantum
 secret sharing (QSS), was proposed by Hillery, Bu\v zek, and Berthiaume
 (HBB99) \cite{12}. Through this scheme a common key or a secret
 message can be securely distributed to many parties simultaneously, who
 are not entirely trusted. Then, while no one alone can recover the message,
 the receivers can cooperatively recover the message by combining the whole
 distributed information of the sender. Since this cryptography scheme of QSS
 can be used effectively when a sender wants to transmit a message to many
 untrusted parties simultaneously, a lot of works have succeeded
 theoretically \cite{13,cgl,15,22} and experimentally \cite{26}
 as one of the most important applications of quantum cryptography.

  Most of the proposed QSS protocols use an entangled state \cite{12,13,15,26,27,28}.
 Yet, the use of the entangled state is not easy for multiparty secret sharing,
 since the efficiency of preparing many-partite entangled state is not high
 enough for real application \cite{y12,y13}. In addition, the communication
 efficiency of the proposed QSS protocols has reached in maxmum 50\%
 in principle \cite{gg}.

  Meanwhile, two QSS protocols without entanglement were proposed, which
 were modifications of BB84 \cite{gg}. One protocol is that a sender creates
 a two-qubit product state in the base $Z = \{ |0 \rangle, |1\rangle \}$ or
 $X = \{ |+ \rangle, |- \rangle\}$ and sends each qubit to the receivers,
 respectively. Then each receiver measures his own received qubit and
 decodes the key bit by coworking with the other receivers. In this protocol,
 the theoretical communication efficiency has doubled in comparison with the
 scheme using an entangled state. Another protocol is that a sender sends
 a string of qubits to receivers, in which key bits are encoded. Each receiver
 shuffles the polarization of the string of qubits with his own parameters and
 sends the string to the next receiver sequentially. Then, the last receiver
 measures the polarization of the string and decodes the information from
 the sender by combining the receivers' shuffling parameters. More recently,
 a single qubit QSS was  proposed experimentally \cite{s14}. However, the
 problem with these protocols is that they are insecure against photon number
 splitting (PNS) attack \cite{nsg} except the case of the use of a single photon
 as a qubit. Moreover, even though the protocols may use a single photon,
 it is not easy to make a reliable single-photon source with current
 technology and photons may be easily lost due to the imperfect channel
 efficiency. To overcome this vulnerability of QSS protocols without
 entanglement, in this paper, we propose a new secure QSS protocol,
 which is without entanglement, and can use a single not-so weak
 coherent laser pulse as a qubit.

\section{Protocol}
  As is shown in the schematic diagram in Fig. 1, our protocol uses a
 randomly polarized qubit. Here the sender Alice sends a single randomly
 polarized not-so-weak pulse to the receiver Rec-$1$. On receiving the
 qubit, Rec-$1$ prepares two rotational angles, an arbitrary random
 angle $\phi_1$ to hide the qubit polarization and the other one
 $s_1\in\{0,\pi/2, \pm\pi/4 \}$ to shuffle the qubit state, and
 rotates the polarization of the received qubit. Then he sends the qubit
 to Rec-$2$. Rec-$2$ puts the received qubit in the same process, but
 with different random angles, $\phi_2$ to hide the qubit polarization and
 $s_2\in\{0,\pi/2, \pm\pi/4 \}$ to shuffle the qubit state, and
 returns the qubit to Alice. Alice compensates her random angle, codes
 a key bit, and randomly shuffles the polarization basis. Then she sends
 the qubit to Rec-$2$. Rec-$2$ compensates his arbitrary random angle
 $\phi_2$ and sends the qubit to Rec-$1$. Rec-$1$ also compensates
 his random angle $\phi_1$ and measures the photon state. When Alice
 publicly announces the basis of the qubit, Rec-$1$ and Rec-$2$
 cooperatively decode and share the key bit. For integrity,
 after error correction and privacy amplification, Alice, Rec-$1$,
 and Rec-$2$ check the shared key bit with a hash function. In this QSS
 protocol each receiver generates his own hidden information so that the
 information given by the sender can be decoded only when the two
 receivers' and the sender's information is joined.

\begin{figure}
 \begin{center}
  \includegraphics[width=8.7cm]{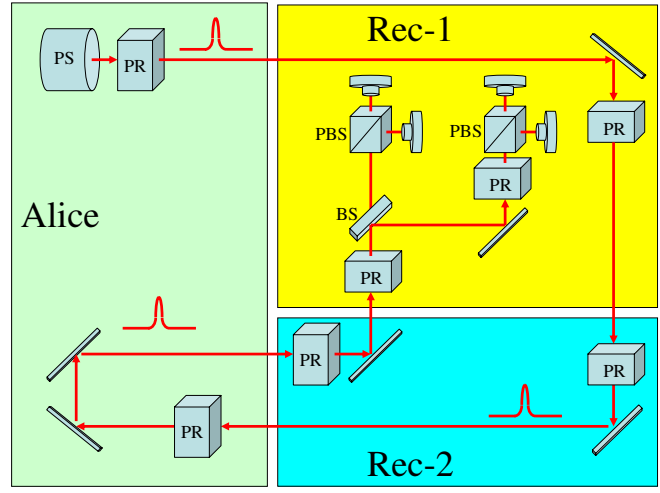}
    \caption{Schematic diagram of QSS protocol with a single qubit. PS is the photon
        source, PR the polarization rotator, BS the beam splitter and PBS the polarizing beam splitter.}
 \end{center}
\end{figure}

{\it Protocol.-} The procedure of the proposed QSS is as follows:
\begin{itemize}

\item [(p.1)] \label{1} The sender, Alice, prepares a linearly polarized
 qubit whose initial state is $|0 \rangle$ and
 prepares an arbitrary random angle $\theta$. Then Alice rotates the
 polarization of the qubit by the prepared angle $\theta$, to bring the state
 $|\psi_0\rangle =\hat{U}_y(\theta)|0\rangle=|\theta \rangle$, where
 $\hat{U}_y(\theta)=\cos\theta\openone-i\sin\theta\hat{\sigma}_y$
 rotating the polarization angle along the \textit{y}-axis. $\hat{\sigma}_y$
 is the Pauli-\textit{y} operator.
 Alice sends the qubit of $|\psi_{0}\rangle$ to Rec-$1$.

\item [(p.2)] \label{2} Rec-$1$ chooses another random angles
$\phi_1$ and $s_1$,
 and rotates the polarizations of the received qubit by $\phi_1+s_1$.
 Then the qubit state is $|\psi_1\rangle = |\theta+ \phi_1+s_1\rangle$.
 Rec-$1$ sends the qubit to Rec-$2$.

\item[(p.3)] On receiving the qubit, Rec-$2$ also chooses other
 random angles $\phi_2$ and $s_2$, and rotates the polarization of
 the received qubit by $\phi_2+s_2$. Then the qubit state becomes
 $|\psi_2\rangle =|\theta+\sum_{i=1}^{2}[ \phi_i+s_i]\rangle$.
 Then  Rec-$2$ sends the qubit to Alice.

\item[(p.4)] Alice rotates the polarization of the received qubit by
 $k_1 \in \{0, \pi/2\}$ or $k_ 2\in \{\pm\pi/4 \}$ to encode the key
 bit, and compensates her random angle by applying $-\theta$. Here, if
 Alice wants to encode '$0$' ('$1$') to the qubit, she rotates the polarization
 of qubit by $\{0, \pi/4\}$ ($\{\pi/2, -\pi/4\}$). The state becomes
 $|\psi_3\rangle = |k_j+\sum_{i=1}^{2}[ \phi_i+s_i]\rangle$,
 where $j \in \{1, 2\}$. Then Alice transmits the encoded qubit to Rec-$2$.

\item[(p.5)] After receiving the qubit, Rec-$2$ compensates his  random
 angle by rotating the polarization of the received qubit by $-\phi_2$, and
 sends the qubit to Rec-$1$.

\item[(p.6)] Rec-$1$ compensates his random angle by rotating the
 polarization by $-\phi_1$, divides the received qubit into two qubits,
 and measures the polarization of each qubit, one on the rectilinear basis,
 $\{|0\rangle, |\pi/2\rangle\}$, and the other on the diagonal basis,
 $\{|+\pi/4\rangle, |-\pi/4\rangle\}$. The reason of the division
 here is because Rec-$1$ does not know the polarization basis on this
 stage. The basis of the qubit is determined by $k_j+\sum_{i=1}^{2}s_i$.
 We can find the basis from Table $1$. So after measurement, Rec-$1$
 stores the results of each measurement until he knows the whole values of
 $k_j+\sum_{i=1}^{2}s_i\equiv l$. Therefore, Rec-1's ultimate decision
 angle is $l-s_1$ and Rec-2's is $s_2$.

\begin{table}
\renewcommand{\arraystretch}{2}
\begin{tabular}{|c|c|c|c|c|}
\hline
\backslashbox{Rec-2}{Rec-1}& ~~~~~~$0$~~~~~~ & ~~~~~$\pi/2$ ~~~~~& ~~~~~$\pi/4$~~~~~& ~~~~$-\pi/4$~~~\\
\hline
0 & 0 & $\pi/2$ & $-\pi/4$ & $\pi/4$ \\
\hline
$\pi/2$ & $\pi/2$ & 0 & $\pi/4$ & $-\pi/4$ \\
\hline
$\pi/4$ & $\pi/4$ & $-\pi/4$ & 0 & $\pi/2$ \\
\hline
$-\pi/4$ & $-\pi/4$ & $\pi/4$ & $\pi/2$ & 0 \\
\hline
\end{tabular}
\label{table1} \caption{Polarization angles from Alice's encoding
 depending on Rec-1's and Rec-2's ultimate decision angles.
 Rec-1 and Rec-2 have a key from mutual relations among their own
 values.}
\end{table}

\item[(p.7)] Here, to make $M$ bits of the key string, the three authorized persons repeat
the above protocol $N (\geq M)$ times. Rec-$1$ can
 lose the key bit for one part of division of some qubits,
 because of the division of the qubit. In other word,
 one part of the division may happen to have a vacuum state.
 If sifted key is shorter than required,
 the parties need only follow the same procedure from the start
 until the sifted key string is of satisfactory length ($M$ bits).
 Rec-$1$ discards $N-M$ qubits which make vacuum states.
 As a result, Rec-$1$ gets $M$ bits of the key string.

\item[(p.8)] Alice publicly announces
 her $M$ basis shuffling factors $j$. However, Rec-$1$ and Rec-$2$ still
 do not know in which basis to look yet, because they know nothing
 about the action of the other. Therefore, the receivers get
 together and compare their polarization angles to recover the key angle string from
 Alice. Then they can find the basis and key angle as shown in Table $1$.

\item[(p.9)] Alice has $M$ bits of $k$, while each of Rec-$1$ and
 Rec-$2$ has $M$ bits of their decision angles, $l-s_1$ and $s_2$, respectively. Rec-$1$
 and Rec-$2$ cooperatively recover the $M$ bits of the key value $k$
 with their own decision angles $l-s_1$ and $s_2$. In this process Rec-$1$
 and Rec-$2$ individually exchange their angles $l-s_1$
 and $s_2$. Then the receivers have the key angles of $k_1$ for
 Rec-$1$ and $k_2$ for Rec-$2$.

\item[(p.10)](Error correction and privacy amplification)
After recovering the key angles, they proceed error correction and privacy
amplification using same method with BB84 protocol.
Alice and receivers can convert the
polarization of key bits into raw key bits regarding a $0$ or $+\pi/4$ as denoting a $'0'$
and a $\pi/2$ or $-\pi/4$ as a $'1'$.

\item[(p.11)](Check of integrity)
  Rec-$1$ and Rec-$2$ publicly announce
 the hash values of the recovered key bit $h_1 = H(k_1)$ and $h_2 = H(k_2)$
 to Alice, respectively, where $H$ is the hash function. Then Alice
 announces her hash value of $h_0=H(k)$ to Rec-$1$ and Rec-$2$, individually.
 They compare the received hash value with their own. When the hash values
 are the same, they share the key bits, otherwise all parties abolish the key
 bits and they repeat the protocol again.
\end{itemize}

  Through this protocol, three parties can share key string through quantum
 channels. Note that, by slightly modifying the procedure of from (p.6) to (p.8)
 in the protocol, we can increase the communication efficiency.

\section{Security}
{\it Honesty ---} As for secret sharing, if one of the receivers is
 dishonest, the dishonest one could get the true information, while
 the others might have wrong one. However, in this protocol, let us assume that Rec-$1$
 knows the true key bits from Rec-$2$'s announcement of his decision
 factor $s_2$. If Rec-$1$ gives wrong information of $l$ or $s_1$ to
 Rec-$2$, Rec-$2$ has wrong information so that $h_1 \neq h_2$.
 Moreover, when Alice compares her hash value with the received
 ones from Rec-$1$ and Rec-$2$, the hash values become
 $h_1 = h_0$ while $h_2 \neq h_0$. Hence Alice knows who is
 dishonest.

{\it Security ---} In QSS, there are various attack strategies depending
 on the protocol \cite{k2}. Since the PNS and the impersonation attacks
 are considered to be the most efficient ones from an eavesdropper's
 prospective, we examine the security of our protocol against those attacks.
 For ease of discussion, we consider that Eve, the eavesdropper, is so
 superior that her action is limited only by the laws of physics. We examine
 the security when a key is encoded into a coherent-state pulse,
 $|\alpha \rangle=|\sqrt{\mu}e^{i\theta}\rangle$, with the mean
 number of photons $\mu$. And the transmission on a line of length
 $\textit{l}$ [km] is $T=10^{-\delta/10}$, where $\delta=\alpha\textit{l}$
 with the losses $\alpha$ [dB/km] on the line \cite{ags}.

{\it PNS attack ---} The conventional strategy of the PNS attack for weak
 coherent laser pulses is that Eve splits off a qubit from each pulse and
 keeps it in the quantum storage. When the basis reconciliation is made
 between Alice and all the receivers via a classical channel, Eve gets the
 secret information by measuring the state of the stored qubit during the
 bases are announced. In this scheme, Eve can measure the number of
 photons with quantum non-demolition method \cite{qnd}. When more
 than two photons are transmitted, she splits off one photon and stores it.
 After the reconciliation procedure between Alice and the receivers Eve
 measures the photon polarization. In our protocol, however, all qubits in
 any step have a completely random polarization. Even if Eve splits off
 some photons from a qubit in the channel between Alice and the
 receivers, she should determine two continuous angles $\theta$ and
 $\phi_i$. Moreover, since Eve does not know the polarization basis,
 she should just estimate the polarization angle with some probability of
 error. If, for simplicity's sake, we assume that each channel has a same
 distance, Eve's mean number of photons of each channel from PNS is
 $\mu T^{c-1}(1-T)$, where $c \in \{1, 3, 4\}$ means $c$-th channel
 of PNS. Then she might try to estimate the polarization with the photons.
 The optimum estimation of the random polarizations in this case can be
 obtained from the fidelity which is calculated to consider the amount of
 information. Depending on the initial amplitude and the transmission rate,
 we find the maximum bound for Eve's information by fidelity as well
 presented in \cite{kcp}. In this paper, for the realistic transmission
 $T=0.5$ and its amplitude $|\sqrt{\mu}e^{i\theta}|=2.83$ give fidelity
 $F_E=0.83$.

   Another possible PNS attack is that Eve counts the number of photons
 at the output of Alice's box. If she finds more than one, she keeps one photon
 and forwards the others. Eve waits until photons come back from Rec-$2$.
 Eve attaches her unpertubed photon to the others, and sends these photons
 to Alice. After the process (p.4), when Alice sends the encoded qubit, Eve
 intercepts the qubit and pick up her photon. At this time, Since Eve knows where
 she attached the photon, she can easily picks up it. Then the polarization
 of Eve's photon is exactly the one chosen by Alice to code her key bit, because
 the photon has not been perturbed by the receivers. Eve keeps the photon and
 sends the others to the receivers. Here, one of the possible ways to pick up
 the same photon is to send her photon in just before or just after the others with
 a very fast switch when photons are transmitted from Rec-$2$ to Alice. When
 the shuffling factors are announced Eve recovers the key bits. This type of attack
 can be prevented easily, when Alice uses a beam splitter. Alice divides her
 photons into two using a beam splitter in the process (p.4) and sends one
 part of the split photons. Since Eve does not know whether her photon is lost or
 not by the beam splitter, Eve gets wrong key bits depending. This counter-attack
 method does not degrade the communication efficiency, because our protocol
 uses a not-so-weak coherent pulse. 

   The second way is that Eve slightly modifies the wavelength of her photon
 through a non-linear process which does not change the polarization. In this
 case Eve also loses her photon because of the beam splitter. If the key length
 is long, Eve can not attack the key bit. For example, when the key length is $256$
 bits and the beam splitter is a 50:50 one, Eve should use a brute text attack for
 the unknown $128$ bit. This is in no ways different from the brute text attack to
 a key of $128$ bit length, which is a usual key length in modern cryptosystem.
 Another counter-attack method is also well explain in Ref.\cite{kk}. According
 to this method, Alice and the receivers can filter out unwanted frequencies by
 using an optical device, which is the combination of an optical grating and
 a very small frequency bandwidth filter like a high resolution Fabry-Perot
 interferometer. In any case, the most simple and efficient way to prevent an
 attack is to use a beam splitter. Hence, the conventional PNS attacks are not
 valid in our protocol as described above even though we can use a
 not-so-weak coherent pulse to make quantum secret sharing.

\begin{figure}
\begin{center}
 \includegraphics[width=8.7cm]{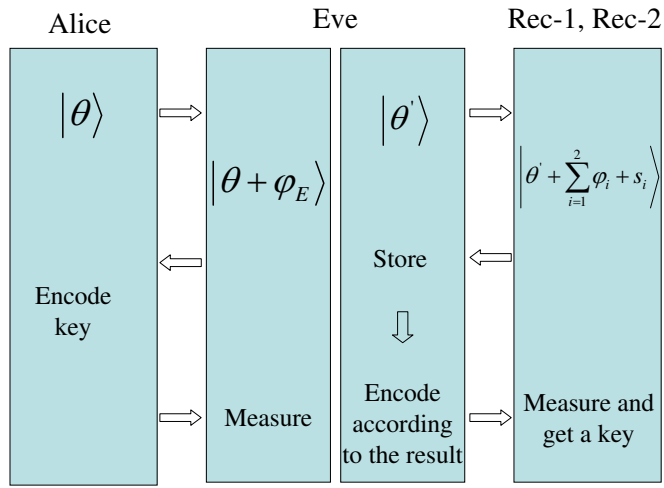}
   \caption{Impersonation attack; Eve impersonates Alice and two receivers, Rec-$1$ and Rec-$2$.}
\end{center}
\end{figure}

{\it Impersonation attack ---} Even if we use a not-so-weak coherent pulse
 instead of a single-photon to make key bits, our protocol is secure against the
 impersonation attack \cite{kcp}, in which Eve impersonates Alice to Rec-$1$
 and Rec-$2$ and vice versa. The strategy of the attack is shown in Fig.2. Eve
 intercepts the transmitting qubit from Alice to Rec-$1$, and sends a different
 pulse which is prepared by herself. When Eve's qubit is returned from
 Rec-$2$, Eve intercepts the qubit and stores it. Eve rotates Alice's qubit and
 sends it to Alice. When Alice sends her qubit to Rec-$2$ after encoding a key,
 Eve intercepts Alice's qubit and measures it after compensating her random
 angle. After the measurement, Eve encodes the key angle, which she has
 measured, to the stored qubit and sends it to Rec-$2$. On this condition, as far
 as Alice does not shuffle the polarization basis, Eve can measure the angle
 without any difficulty. Because of Alice's basis shuffling, however, Eve can not
 encode the key angle perfectly. Let us calculate the error rate of Rec-1's
 measurement. As far as polarization is concerned, Eve has an exact copy of
 Alice's key. Therefore, Eve can obtain full information using strategies based
 on unambiguous state discrimination. Unambiguous discrimination among
 four states of a two-dimensional Hilbert space is only possible when at least
 three copies of the state are available. And the optimal probability for
 discriminating between the four states with $n \geq 3$ copies is found to be
 $p_{ok}(n)=1-(1/2)^{[(n-1)/2]}$, where $[\cdot]$ is the rounding to the
 closest lower integer \cite{ags}. The probability of there being $n$-photons
 in the coherent state with mean photon number $\mu T$, which is reduced
 from $\mu$, is $p_{c}(n, \mu T)=e^{-\mu T}(\mu T)^{n}/n!$. Then the
 maximal probability that Eve can discriminate the state without error is
\begin{eqnarray}
   P_{E}&=&\sum_{n \geq 3}p_{ok}(n) p_{c}(n, \mu T)  \nonumber\\
         &=&\sum_{n \geq 3}\frac{e^{-\mu T}(\mu
       T)^n}{n!}\left\{1-\left(\frac{1}{2}\right)^{[\frac{n-1}{2}]}\right\}.
\end{eqnarray}
 Therefore, the error rate of Rec-1's side is
\begin{eqnarray}
 P_{error}=\frac{1}{2}(1-P_E)
\end{eqnarray}
 since Eve's wrong encoding makes $1/2$ error rate on Rec-1's side.
 The error rate depending on the mean photon number is plotted in
 Fig. 3. If the error rate from his measurement is similar
   to $P_{error}$, he can ascertain whether Eve exists or not.
   The figure shows that for $T=0.5$ and $\mu \simeq 6$ the error
 rate is $P_{error}=0.3$. In other words, when Alice sends a coherent state
 pulse with $\mu=6$, the criterion to check eavesdropper is
 $P_{error}=0.3$.

\begin{figure}
\begin{center}
 \includegraphics[width=9.6cm]{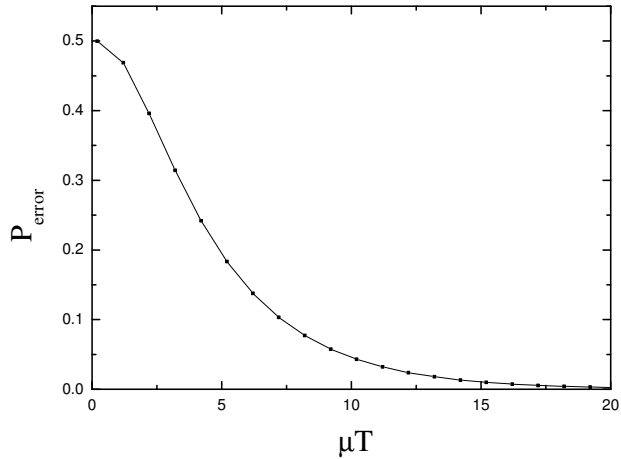}
   \caption{Rec-1's error rate $P_{error}$ depending on Eve's optimal probability
   for discriminating between the four states under the number of $n$
   photons $\mu T$; when Rec-1 knows mean photon number, he gets error rate to check
   the existence of Eve.}
\end{center}
\end{figure}


\section{General Protocol}

 In QSS, a sender usually shares a message with more than two receivers
 simultaneously.
 In our protocol, since we can use a single not-so-weak coherent pulse, it is
 easy to generalize the QSS protocol for the case of $(N+1)$ multi-parties
 ($N \geq 2$); one is a sender Alice and the others receivers.

{\it General protocol.---}The general protocol among $(N+1)$ parties
is as follows:
\begin{itemize}

\item [(F.0)] \label{1} Alice prepares a randomly polarized qubit
 $|\psi_0\rangle=|\theta\rangle$ and sends it to Rec-$1$.

\item [(F.1)] \label{2} Rec-$1$ chooses his random angle $\phi_1$
 and rotates the polarizations of the received qubit by $\phi_1+s_1$,
 where $s_1\in \{0, \pm\pi/4, \pi/2\}$. Rec-$1$ sends the qubit
 to Rec-$2$.

\item [(F.n)] \label{3} In this way, Rec-$n$ $(n \in\{2,...N-1\})$
 receives the qubit and repeats the same polarization procedure as
 his precedent. Then he sends it to Rec-$(n+1)$.

\item [(F.N)] \label{4} The state of the last receiver Rec-$N$
 becomes $|\psi_N\rangle=|\theta+\sum_{i=1}^{N}[\phi^i+s_i]
 \rangle$. Then, Rec-$N$ sends it to Alice.

\item [(F.N+1)] \label{5} After receiving the qubit, Alice compensates
 her private angle with $-\theta$ and encodes her key angle, $k$. Then
 the state becomes $|k+\sum_{i=1}^{N}[\phi^i+s_i]\rangle$. Alice
 sends it back to \textit{N}$^{th}$ party.

\item [(B.N)] \label{6} Rec-$N$ receives the qubit from Alice and
 rotates the polarization angle with $-\phi_N$. The state becomes
 $|k+\sum_{i=1}^{N-1}[\phi^i]+\sum_{i=1}^{N}[s_i]\rangle$.
 Then he transmits the qubit to Rec-$(N-1)$.

\item [(B.n)] \label{7} In this way, Rec-$n$ receives the qubit from
 Rec-$(n+1)$. After rotating the polarization with Rec-$n$' angle. Then
 the state becomes $|k+\sum_{i=1}^{n-1}[\phi^i]+\sum_{i=1}^{N}[s_i]\rangle$.

 \item [(B.1)] \label{8} Then, Rec-$1$'s qubit state after compensating
 his random angle becomes$|k+\sum_{i=1}^{N}[s_i]\rangle$.
 Rec-$1$ measures the polarization in the same way with (p.6).

\item [(B.0)] \label{9} Since each party has his own decision angles
 $s_i$, they recover the key bits cooperatively. Then they can share
 the key from Alice. By repeating this procedure similar to the protocol between three parties, all receivers
 get the $M$ bits of key string.
\end{itemize}

Similar to the $N=2$ case, Alice and all the receivers check the
shared key bit with hash functions for the integrity. After
receiving the key bits, all receivers publicly announce their hash
values of the recovered key to Alice independently. Alice announces
her hash value to all receivers, simultaneously. Then they compare
the received hash value with their own $H(k)=H(k_n)$ where $k$ is
Alice's key bit and $k_n$ is the $n^{th}$ receiver's key bit. When
all hash values are the same, they share the key bit, otherwise the
key values are abolished and they repeat the protocol again.

\section{Discussion and Summary}
  Our QSS protocol has several advantages in comparison with the other QSS
 schemes:

 (1) Even though our protocol uses not-so-weak coherent pulses,
 the random polarization and the basis shuffling can effectively block up the
 PNS and the impersonation attacks, while the other QSS protocols without
 entanglement are vulnerable when they use not-so-weak coherent pulses.
 So our QSS protocol is easier to realize experimentally with modern technology
 than the other QSS protocols. We have modified the BB84 protocol to achieve stronger QSS protocol.
 Similarly, we can also make the QKD protocol stronger than BB84 for the
 above reasons.

 (2) In the step of (p.6), Rec-$1$ splits the received qubit into two parts to measure
 the polarization with two different bases. To split the qubit, Rec-$1$ should get
 coherent-state pulse with mean photon number at least $2$. As shown in Fig. 3,
 we can use multi-photon pulse and control the mean photon number by referring the
 relation between $P_{error}$ and mean photon number. Therefore, comparing with the
 other QSS protocols without entanglement, we do not need to employ quantum memory to
 store the qubit even if we use non-orthogonal bases.

 (3) In this protocol, if one of the receivers, Rec-$i$, is dishonest and
 sends wrong decision factor to the other receivers, each of the honest receivers generates
 a different hash value. Alice and honest receivers can easily notice that Rec-$i$ is
 dishonest by checking the integrity with hash values. Therefore, we can
 discern dishonest behavior among the receivers.

 (4) Even though our protocol
 uses a classical channel to distribute the information of the measurement basis from
 Alice, neither Eve nor any receiver can obtain the information from Alice,
 because all receivers shuffle the polarization bases. In other words, the measurement
 basis of each receiver depends on the shuffling of the other receivers. This is another
 reason why our protocol is robust against the PNS attack. Consequently,
 compared with other protocols, communication through the classical channel of our protocol reveals no
 information to Eve or any dishonest receivers.

  In summary, we have proposed a new protocol of quantum secret sharing
 with a single not-so weak coherent pulse for multipartite receivers. In this protocol,
 because each receiver shuffles the polarization bases, no one alone can
 obtain the secret key transmitted by a sender, while all receivers can
 recover the key cooperatively. A dishonest receiver can be noticed by
 exchanging the hash values of the recovered key between communicators.
 Finally, since all of the operations are random and independent, the
 security is guaranteed against the PNS and the impersonation attacks.

\section*{Acknowledgements}

  This work is supported by Creative Research Initiatives of the Korea
 Ministry of Science and Technology. Y.J. Choi and Y.-J. Park are
 supported by the SRC Program of the Korea Science and Engineering
 Foundation (R11-2005-021). J. Kim was financially supported by Korea Ministry of Information and Communication
  under the "Next Generation Security" project.


\begin{thebibliography}{9}

\bibitem{k1} Bennett C H and Brassard G 1984 {\it Proc. IEEE
 Int. Conf. Computers, Systems, and Signal
 Processing (Bangalore)} (New York:IEEE) p 175

\bibitem{k4} Ekert A K 1991 {\it Phys. Rev. Lett.} {\bf 67} 661

\bibitem{k2} For a review, see Gisin N, Ribordy G, Tittel W, and
 Zbinden H 2002 {\it Rev. Mod. Phys.} {\bf 74} 145

\bibitem{k3} Curty M, Lewenstein M, and Lutkenhaus N 2004 {\it Phys. Rev.
Lett.} {\bf 92} 217903

\bibitem{12} Hillery M, Bu\v zek V and Berthiaume A 1999 {\it Phys. Rev.
A} {\bf 59} 1829

\bibitem{13} Karlsson A, Koashi M and Imoto N 1999 {\it Phys. Rev. A} {\bf 59} 162

\bibitem{cgl} Cleve R, Gottesman D and Lo H-K 1999 {\it Phys. Rev. Lett.} {\bf 83} 648


\bibitem{15} Karimipour V, Bahraminasab A and Bagherinezhad S 2002 {\it Phys.
 Rev. A} {\bf 65} 042320

\bibitem{22} Zhang Z J, Li Y and Man Z X 2005 {\it Phys. Rev. A} {\bf 71} 044301

\bibitem{26} Tittel W, Zbinden H and Gisin N 2001 {\it Phys. Rev. A} {\bf 63} 042301

\bibitem{27} Lance A M, Symul T, Bowen W P, Tyc T, Sanders B C and Lam P K 2003 {\it New J. Phys.} {\bf 5} 4

\bibitem{28} Li X H, Zhou P, Li C-Y, Zhou H-Y and Deng F-G 2006
{\it J. Phys. B: At. Mol. Opt.} {\bf 39} 1975

\bibitem{y12} Bouwmeester D, Pan J W, Daniell M, Weinfurter H and Zeilinger A 1999 {\it Phys. Rev. Lett.} {\bf 82} 1345

\bibitem{y13} Pan W, Daniell M, Gasparoni S, Weihs G and Zeilinger A
2001 {\it Phys. Rev. Lett.} {\bf 86} 4435

\bibitem{gg} Guo G P and Guo G C 2003 {\it Phys. Lett. A} {\bf 310} 247
; Yan F-L and Gao T 2005 {\it Phys. Rev. A} {\bf 72} 012304

\bibitem{s14} Schmid C, Trojek P, Bourennane M, Kurtsiefer C, Zukowski M and Weinfurter H
2005 {\it Phys. Rev. Lett.} {\bf 95} 230505

\bibitem{nsg} Niederberger A, Scarani V and Gisin N 2005 {\it Phys. Rev. A} {\bf 71} 042316

\bibitem{ags} Ac\'{\i}n A, Gisin N and Scarani V 2004 {\it Phys. Rev.
A} {\bf 69} 012309

\bibitem{qnd} Naik D S {\it et al.} 2000 {\it Phys. Rev. Lett.} {\bf 84} 4733
; Pryde G J, O'Brien J L, White A G, Bartlett S D and Ralph T C 2004
{\it Phys. Rev. Lett.} {\bf 92} 190402

\bibitem{kcp} Kye W-H, Kim C-M, Kim M S and Park Y-J 2005
{\it Phys. Rev. Lett.} {\bf 95} 040501

\bibitem{kk} Kye W-H and Kim M S quant-ph/0508028





\end{thebibliography}
\end{document}